\documentclass[11pt, a4paper, onecolumn, copyright, goog]{google}

\usepackage[authoryear, sort&compress, round]{natbib}
\keywords{deliberation, large language models, artificial intelligence}
\paperurl{https://arxiv.org/abs/123}

\uselogo{} 

\title{Can AI Mediation Improve Democratic Deliberation?}

\correspondingauthor{mhtessler@google.com, georgieevans@google.com}

\usepackage{hyperref}

\defcitealias{tessler2024ai}{Tessler, Bakker, et al.}
\newcommand{\citetessler}{\citetalias{tessler2024ai}~(\citeyear{tessler2024ai})}

\newcommand{\citetesslerp}{(\citetalias{tessler2024ai}, \citeyear{tessler2024ai})}

\reportnumber{} 

\renewcommand{\today}{2025-06-09}

\author[*,1]{Michael Henry Tessler}
\author[*,1]{Georgina Evans}
\author[1,2]{Michiel A. Bakker}
\author[1]{Iason Gabriel}
\author[1]{Sophie Bridgers}
\author[1]{Rishub Jain}
\author[1]{Raphael Koster}
\author[1]{Verena Rieser}
\author[1]{Anca Dragan}
\author[1,3]{Matthew Botvinick}
\author[1,4]{Christopher Summerfield}

\affil[*]{Equal contribution}
\affil[1]{Google DeepMind, London, UK}
\affil[2]{Massachusetts Institute of Technology, MA, USA}
\affil[3]{Yale Law School, New Haven, CT, USA}
\affil[4]{Department of Experimental Psychology, University of Oxford, Oxford, UK}

\begin{abstract}
The strength of democracy lies in the free and equal exchange of diverse viewpoints. Living up to this ideal at scale faces inherent tensions: broad participation, meaningful deliberation, and political equality often trade off with one another \citep{fishkin2011trilemma}. We ask whether and how artificial intelligence (AI) could help navigate this “trilemma” by engaging with a recent example of a large language model (LLM)-based system designed to help people with diverse viewpoints find common ground \citetesslerp. Here, we explore the implications of the introduction of LLMs into deliberation augmentation tools, examining their potential to enhance participation through scalability, improve political equality via fair mediation, and foster meaningful deliberation by, for example, surfacing trustworthy information. We also point to key challenges that remain. Ultimately, a range of empirical, technical, and theoretical advancements are needed to fully realize the promise of AI-mediated deliberation for enhancing citizen engagement and strengthening democratic deliberation.
\end{abstract}

\begin{document}

\maketitle
\footnote{This paper was presented at the Knight Institute for the First Amendment at Columbia University workshop on ``AI and Democratic Freedoms'', April 10-11, 2025, and is available on the Knight Institute website \url{knightcolumbia.org/}.}

\section{Introduction}

In the early 2000s, concern arose in the United States about the fairness and efficacy of the allocation process for kidney transplants. The system at the time pursued a seemingly robust objective: namely, it respected the time spent on the waiting list and the expected number of additional years of life for the recipient. However, there were valid objections: Did it discriminate unfairly against the elderly by virtue of them leading healthy lives? Did it recognize that the medical urgency of situations differed? Did it treat different demographic groups fairly? The experience of people placed on the waiting list suggested not. To address these concerns, public consultations took place in which everyone affected was encouraged to share their viewpoint and engage in deliberation. After numerous rounds of feedback, a new algorithm was created that weighs ten different variables, a change widely praised by medical practitioners and patient groups alike. Commenting on the process, scholar David Robinson notes that “[E]xperts, patients, and advocates balanced hard trade-offs to remake the kidney allocation algorithm, leading to better health for more people and a fairer allocation of organs…People raised their voices. People were heard” \citep{robinson2022voices}. The only serious limitation was that the process took 20 years to complete.

Imagine if we could channel the power of public deliberation effectively and efficiently, providing timely guidance on a range of issues. Traditional methods for public consultation have well-known limitations. Focus groups can provide detailed feedback but are too small scale to be representative and are susceptible to biases affecting group discussion \citep{macdougall1997devil, o2018use}. Canvassing opinions door-to-door to elicit large-scale input can be costly and time-consuming. Running a poll can help reveal the balance of opinion but typically is restricted to a small number of judgments, does not allow for discussions of different points of view, and can be sensitive to the wording of the questions \citep{gallup1941question, desaint2013use}. More structured forms of public input, like deliberative polls or citizens' assemblies, involve meaningful deliberation but also require significant time commitments from participants \citep{fishkin2003consulting}. Consequently, well-resourced groups—who have more time, better coordination, or greater knowledge about the process—can dominate and be over-represented in these settings \citep{birhane2022power}, and these offline methods are not scalable to very large groups of people.

Recent efforts to overcome the scalability and logistical challenges of traditional consultation methods take shape in digital deliberation technologies \citep{coleman2012connecting, goni2025citizen}. Systems like Pol.is and Remesh facilitate large-scale input by allowing participants to contribute statements and react to others, typically through voting \citep{small2021polis}. These platforms process the interaction data (e.g., votes) with machine learning techniques in order to uncover collective sentiments by grouping participants based on shared opinions and highlighting points of consensus or division. Despite the significant expansion of scale of participation and efficient ways to map general agreement, these tools are limited in their capacity to process the rich semantic content of participant contributions. Further, the deliberation they support is structured around discrete statements and voting and may miss out on deeper engagement with the reasoning and substance of diverse perspectives, critical for finding agreement and consensus in a population \citep{habermas1985theory}.

Recent advances in artificial intelligence (AI), specifically the emergence of increasingly capable large language models (LLMs), have introduced new opportunities to further support deliberative and democratic functions \citep{landemore2023can, small2023opportunities, lazar2024can}. LLMs can process vast amounts of text and can be prompted or trained further to perform arbitrary tasks involving text \citep{brown2020language}.\footnote{More broadly, large multimodal models and advanced AI agents can process text, images, videos, and sound, among other modalities, and take actions (e.g., on a computer). We focus our discussion on text-based large language models.} Researchers and practitioners have explored how LLMs can support democratic deliberation by summarizing content \citep{small2023opportunities}, aggregating opinions in a manner to increase collective support \citep{fish2023generative}, representing individual judgments in decision-making processes \citep{jarrett2023language, gudino2024large}, facilitating public deliberation \citep{ma2025towards}, and implementing principles derived from democratic methods \citep{huang2024collective}. Despite lots of speculation and demonstrations of the strengths and shortcomings of LLMs to support various deliberative tasks, there has been relatively little empirical work evaluating their efficacy.

Recently, \citetessler\space built and investigated an LLM system—which they call the Habermas Machine (HM)—designed for the task of finding common ground between people with different viewpoints (Figure 1). The AI system produced high-quality statements expressing common ground garnering levels of endorsement from the participants exceeding that of human mediators (in addition, the HM performed this task in seconds, in contrast to human mediators, who required several minutes). The HM works by generating a (potentially large) set of possible group statements and then returning one (or a few) of them by simulating an election in which each participant's ranking of the statements is predicted. This paper provides empirical support for the fairness and efficacy of an AI-powered deliberative technology—the HM—that in principle could mediate deliberation among very large groups. Is fair, time-efficient mass deliberation within reach?

The prospect of AI mediation improving democratic deliberation raises the question of how it might navigate Fishkin’s trilemma—the inherent trade-offs between broad participation, meaningful deliberation, and equality of contribution \citep{fishkin1991democracy}. How might AI mediation enable scaling deliberation? How can we assess whether AI mediation can be fair to all participants? Can AI enhance the quality of deliberation itself, e.g., by aiding access to trustworthy information? This paper explores these key questions, examining the technical foundations of AI-mediated deliberation through the lens of the HM. We additionally analyze the steps needed to bridge the gap between current capabilities and practical application in public decision-making.

In this paper, we focus on a specific type of AI mediation in deliberation, one centered on bridging communication between participants. The AI mediator's primary role, as exemplified by the HM, is to synthesize individual viewpoints and generate statements that reflect common ground and are intended for approval by the involved parties. Beyond the scope of our discussion is the process of gathering opinions and the task of real-time discourse management (e.g., ensuring equitable speaking time or moderating the flow of conversation; \citealt{kim2020bot, shin2022chatbots, ma2025towards}). These aspects of deliberation, while vital, are assumed to occur outside the purview of this particular AI mediator, perhaps facilitated by separate AI systems, human moderators, or integrated within a broader deliberative design.

\begin{figure}[htbp]
    \centering
    \fbox{\includegraphics[width=0.3\textwidth,
                trim=0pt 0pt 8pt 0pt, 
                 clip 
                ]{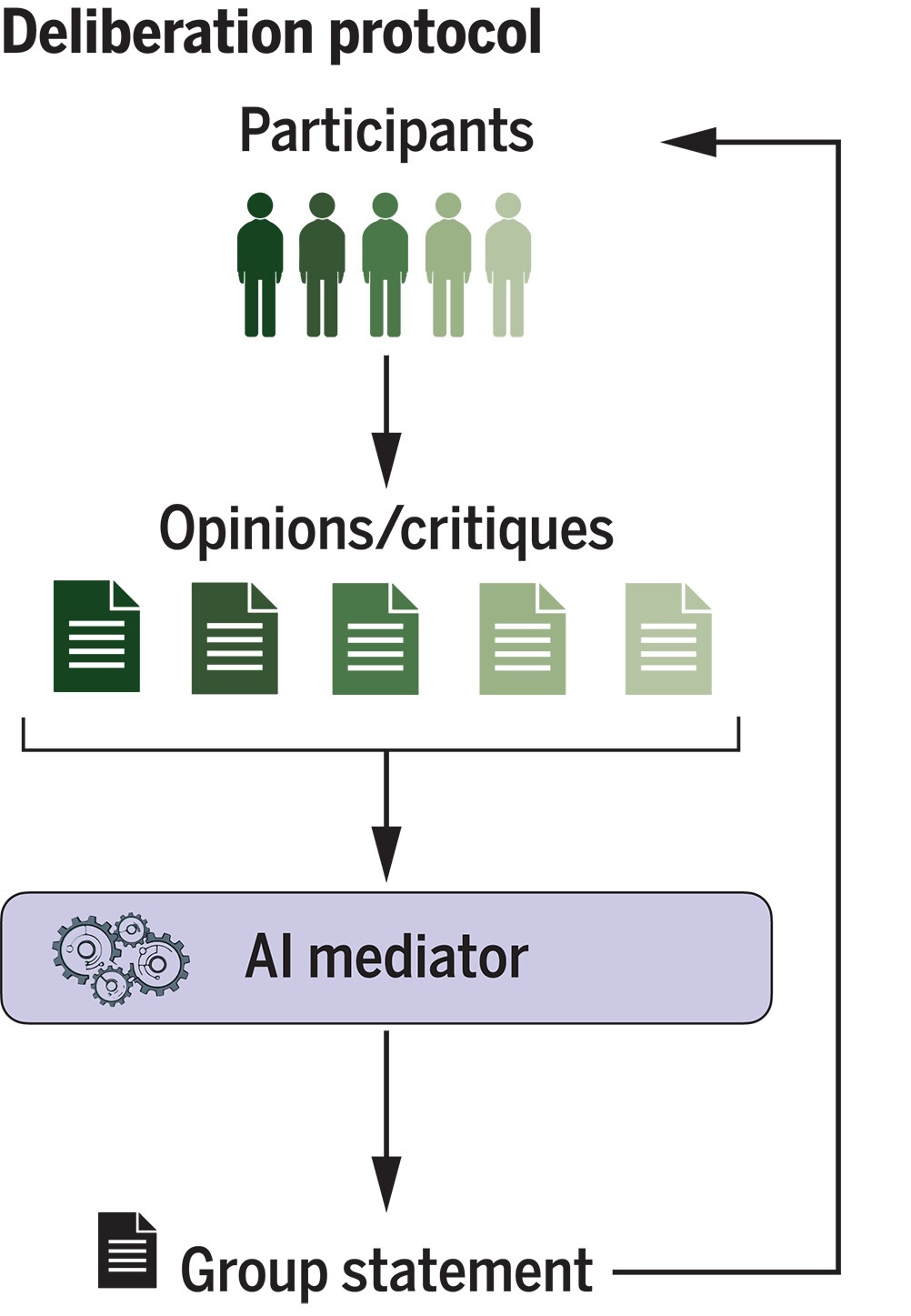}}
    \caption{Schematic overview of AI-mediated deliberation procedure studied in \citetessler. Participants write their opinions, which are sent to the mediator to craft a Group Statement. Participants may then write critiques of the Group Statement, which the AI mediator could then revise. From \citetessler. Reprinted with permission from AAAS.}
    \label{fig:deliberation_schematic}
\end{figure}

\section{AI in Deliberation: The Habermas Machine}

Introducing new technology into deliberation can be seen as an attempt at an answer to Fishkin’s “trilemma” \citep{fishkin2011trilemma}. Fishkin’s trilemma highlights a fundamental tension between three democratic goals: political equality (everyone’s voice or vote is counted equally), inclusion (as many people as possible exercise their voice or right to vote), and deliberation (citizens thoughtfully consider issues). Large-scale voting captures equality and inclusion, but not deliberation; online discussion fora may have deliberation and inclusion, but unequal access or organized interests can lead to a lack of equality; structured deliberation exercises (such as Stanford’s Deliberative Polls) prioritize equality of voice and deliberation but are difficult to scale to large groups. Addressing these trade-offs is therefore a key driver behind the exploration of technological solutions, and especially those involving AI \citep{landemore2023can}.

The field of AI-enabled deliberation is growing quickly \citep{kim2020bot, shin2022chatbots, argyle2023leveraging, michael2023debate, small2023opportunities, ma2025towards}. In order to understand the potential of AI in this field we focus on one particular model—the Habermas Machine—to ground the discussion in a specific example that has been evaluated for its efficacy in experimental settings \citetesslerp.\footnote{Jürgen Habermas is a political theorist who emphasized the importance of communication in reaching mutual understanding and consensus \citep{habermas1985theory}.} The HM also serves as a useful example of a large language model technology being inserted into human deliberation, raising technical and ethical questions. 

\subsection{Deliberation protocol}

The HM, as described in \citetessler, is a system of LLMs that was designed with the goal of finding common ground among people with diverse perspectives (Figure 2B). The HM is an LLM system which handles a list of inputs, coinciding with the opinions, critiques, or comments from a group of individuals. Interaction with the HM begins by group members submitting their personal opinions on a topic (e.g., in response to ‘Should we lower the voting age to 16?’) as a text statement (e.g., a short paragraph outlining their position and justification). The HM then generates a set of ‘group opinion statement(s)’—statements that try to reflect the common ground among the group of individual opinion writers. Participants then vote on which one they prefer. The group members then are invited to write comments or critiques on the (winning) statement (e.g., expressing approval or disapproval, ways to improve the statement, etc.). These critiques are sent to the HM, which then generates a new set of revised group opinion statements. Another round of voting occurs, and the process could continue for more rounds of critique and revision. In \citetessler, the process ended after two rounds (i.e., opinions and one round of critiques). 

\subsection{The HM technical details}

The HM has two LLM components: a generative model and a reward model. The generative model proposes a number of candidates (e.g., 32) for a ‘group opinion statement’ based on the individual opinion statements submitted. These statements explore different wordings and ways of combining the opinions, ideally spanning a diverse set of possible group opinions. The reward model predicts how much each group member would agree with each of the generative model’s candidates, based on the group member’s individual opinion (i.e., would the person with opinion X like statement Y?).\footnote{In \citetessler, the reward model was fine-tuned using the voting (or, ranking) data provided when participants ranked multiple candidates. The generative model was fine-tuned with statements that were rated as high quality from participants.} The reward model predictions are converted into rankings such that each group member has a predicted ranking over the candidate statements. The rankings are aggregated using social choice theory (e.g., the Schulze method, a way of implementing a ranked-choice voting scheme), and the winning candidate is shown to the group members (or some number of the top candidates could be shown). The HM also has the capacity to revise a group statement by incorporating written critiques from the individual group members through the same generation and selection process.

In other words, the Habermas Machine implements a simulated election. Candidates are proposed from a generative model, and votes are determined through a reward model. Votes (rankings) are aggregated in a way that the outcome of a ranked choice election would be determined, and the winning statement is returned to participants for them to write comments or critiques. The process iterates until some predetermined stopping point (e.g., after a certain number of rounds of critiques).

The LLM components of the HM were fine-tuned through subsequent rounds of data collection in the deliberation protocol shown in Figure 2A. The system began as a ‘pretrained’ LLM, meaning that its initial state was as a next-word prediction machine. The authors then further trained two instances of this LLM into a generative model and a reward model, through feedback collected during the deliberation protocol. One could also build an HM out of a system of prompted LLM-based chatbots (e.g., Gemini, ChatGPT, Claude) as well, which would have some benefits and drawbacks (discussed below).

\begin{figure}[htbp]
    \centering
    \fbox{\includegraphics[width=\textwidth]{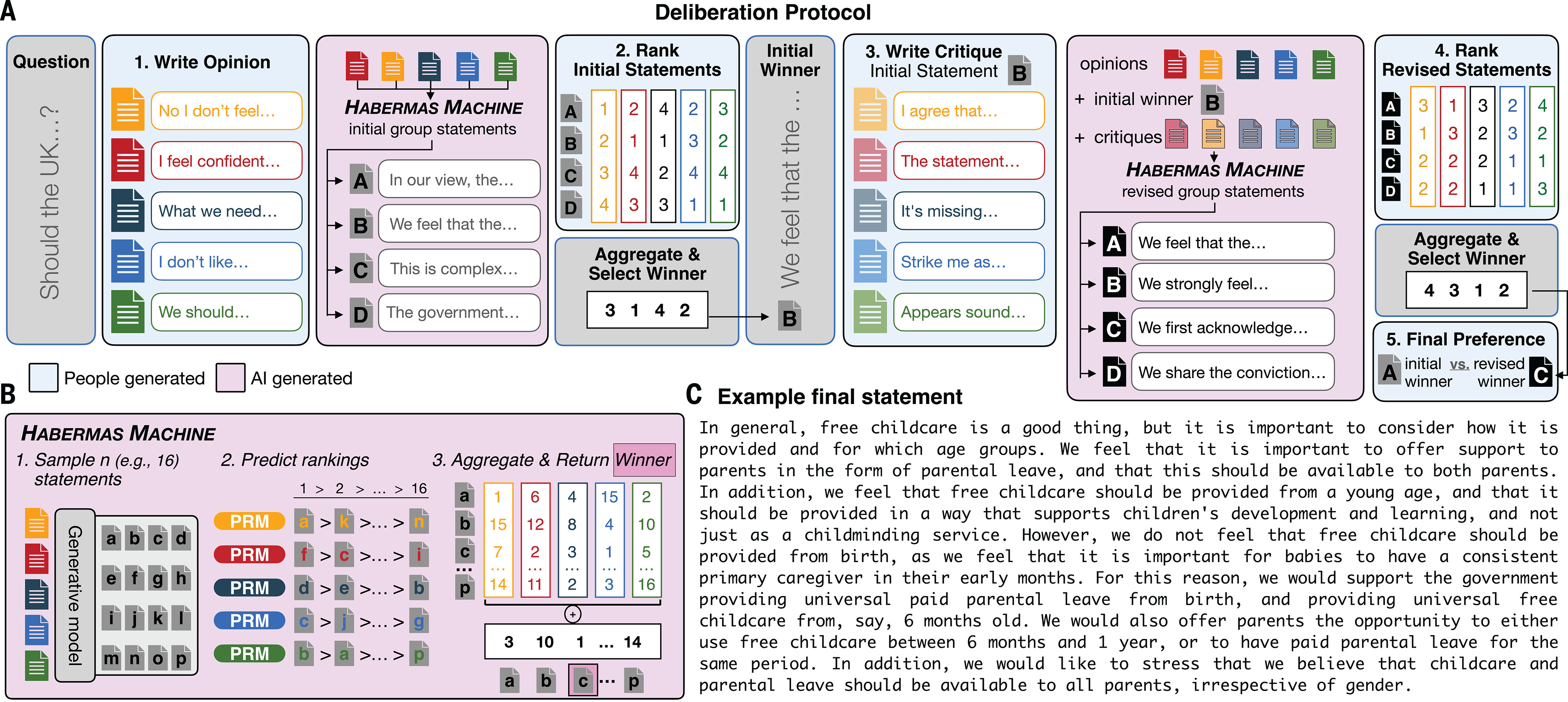}}
    \caption{(A) Full deliberation procedure. 1. Participants, organized into small groups, composed opinion statements in response to a specific question. The Habermas Machine (HM) then generated candidate initial group statements based on these individual opinions. 2. Participants ranked these initial statements. The statement with the highest aggregated ranking was returned to the group. 3. Participants wrote critiques of the initial winning statement. Using these critiques, along with the initial opinions and the initial group winner, the HM generated revised group statements. 4. Participants ranked these revised statements, and the top-ranked statement was selected through aggregated rankings. 5. Participants made a final preference judgment between the initial and revised winning statements. Each deliberation round for a single question lasted approximately 15 minutes. (B) The HM produces a group statement through a simulated election. 1. A generative model samples multiple candidate group statements. 2. A personalized reward model predicts rankings for each person in the group. 3. The statement with the highest aggregated ranking is returned. C) Example winning revised group opinion statement. From \citetessler. Reprinted with permission from AAAS.}
    \label{fig:deliberation_schematic}
\end{figure}

\subsection{Summary of findings}

The authors found that AI mediation indeed helps participants find common ground, outperforming human mediation. The authors note, however, that “the key translational opportunity provided by the HM is not its potentially ‘superhuman’ mediation but rather its ability to facilitate collective deliberation that is time-efficient, fair, and scalable.” Additionally, multiple rounds of deliberation (e.g., having participants write critiques of the HM statements) led to HM statements that received more approval from the participants, even controlling for the fact that they spent more time and effort to arrive at those revised statements. The AI-mediated deliberation also left groups less divided on the issue than they were prior to deliberation, indicating that some amount of belief change occurred from the deliberation. The Habermas Machine was found to fairly represent majority and minority viewpoints (i.e., weighing their opinions in proportion to the size of the majority/minority) after processing the opinions, but after processing the critiques, the HM tended to over-weight minority viewpoints. Finally, the authors replicated their findings in a demographically representative sample of the UK, finding convergent shifts in position across groups on certain topics. An example final group statement is shown in Figure 2C, and many examples can be found in the Supplementary Materials (SM 6) of \citetessler. The deliberation questions and the data collected are publicly available at: \url{github.com/google-deepmind/habermas_machine}.

\section{Addressing the Deliberative Trilemma}

\citetessler\space  offer a noteworthy experimental demonstration that AI-mediated deliberation can help diverse small groups find common ground on potentially divisive topics. These results naturally raise the possibility that AI mediation could resolve the tension between mass participation, equality of contribution, and quality of deliberation that \cite{fishkin1991democracy} introduced. While the trilemma forms the core of the following analysis, we will explore the broader implications, including the role of AI in the democratic process and its comparison to traditional deliberation, in the Discussion.

\subsection{Equality: How can we ensure an AI mediator is fair?}

Numerous principles can contribute to a fair democratic process, such as transparency, inclusivity, equality in decision-making, and safeguarding minority rights against majority rule. While a normative discussion of which principles are essential is beyond scope, it is worth asking whether an AI system could uphold any reasonable interpretation of democratic fairness \citep{barocas2023fairness}. How can we design and implement AI mediators like the Habermas Machine to ensure they operate in a manner that is demonstrably fair and equitable to all participants?

The fact that language models do not adhere to prescribed, deterministic rules in generating outputs is precisely what empowers them to effectively handle a vast and diverse array of inputs. But this opacity and unpredictability can also undermine the legitimacy of the system’s output, clashing with the democratic norms of transparency and mutual understanding of the process. How can we ensure that the statements generated by the Habermas Machine, or a similar technology, are derived fairly and faithfully from participants, and that participants can trust the system to do so? We argue that principles of democratic fairness need to be explicitly incorporated into the design of the system, and then combined with interpretability tools that continuously verify and monitor that the AI mediator is aligned with those principles.  

\subsubsection{Building fairness into the architectural design of an AI mediator}

Part of what differentiates the Habermas Machine from an out-of-the-box (prompted) large language model assistant (such as Gemini, Claude, ChatGPT, etc.) is that the HM produces a statement by simulating an election among a large number of candidate statements. That is, the HM generates an arbitrary number of ‘candidates’ that it considers presenting and uses a “personalized reward model” (PRM) to predict the rankings assigned to the candidates by each person in the group. These rankings then get aggregated by a social choice function (i.e., a ranked choice voting process) to decide upon the statement(s) to return to participants for their consideration. Through ablation analyses, \citetessler\space found that the simulated election (i.e., PRM + aggregation) part of the system was most responsible for producing the statements that the groups of people endorsed highly.  

The HM architecture has three core components: a generative model for sampling candidate group statements, a PRM to rank these statements according to the predicted preferences of each participant, and an aggregation mechanism to choose a final statement based on these PRM-predicted rankings. We discuss how fairness can be built into each component. 

\paragraph{Fair aggregation}

Aggregation is the process of combining a set of individual preferences into a collective ranking. It is the process of counting votes in an election. Both social welfare and social choice frameworks could be used for aggregating preferences over a set of candidates. In a predecessor of the Habermas Machine, \cite{bakker2022fine} explored different social welfare functions ranging from a utilitarian welfare function (weighing all participants’ preferences equally) to a Rawlsian welfare function (only taking the most dissenting participant into account). While this social welfare framework is computationally simpler, this approach relies on the questionable assumption that the strength of one person's preference can be directly weighed against another's.  This may make the system more susceptible to participants who might try to gain an advantage by dramatically overstating how much they like or dislike an option (cf., \citealt{parkes2012complexity}).

The incorporation of this well-defined aggregation mechanism—one that takes a set of participant rankings over candidates and then outputs a final winning group statement—provides a transparent lever to control how different perspectives are represented and weighed in formulating the deliberative output. In principle, the general architecture can adapt to different normative notions of fairness. For instance, by adjusting the social choice aggregation mechanism, one could explicitly amplify the voices of the minority. We discuss the social choice framework and the alternative of social welfare aggregation later in this section.

\paragraph{Fair ranking}

Inside an AI mediator like the HM, there is a PRM that should fairly and faithfully produce personal rankings over statements (i.e., predict how each person in the group would rank a set of hypothetical statements). If aggregation is the process of counting votes, ranking is the process of voting. Fairness requires that the model should be equally accurate for everyone, and faithfulness requires a baseline level of functionality for everyone. If some voters have a difficult time reporting their votes (e.g., with a confusing ballot), the outcome might be unjust. Similarly, if the model is more likely to misinterpret the preferences of certain individuals, for example, those from underrepresented groups or those who struggle to articulate their views, the aggregated outcome may unfairly reflect the preferences of the better-represented or more articulate participants. This problem is salient as large language models may be less accurate in predicting preferences of those who communicate most naturally in non-standard English or other languages \citep{hofmann2024ai, fleisig2024linguistic}.

Adjusting the aggregation mechanism (fair aggregation) can help correct for some misspecification of preferences. For example, instead of forcing the model to make a single, best guess about how a person feels (a "pointwise estimate"), the system could consider a range of possibilities and its confidence in each one (an "expected utility"). This allows the system to account for the fact that it might be more uncertain about some participants' preferences than others \citep{azari2014statistical}. Similarly, "robust aggregation" uses methods that are less easily swayed by  outliers or errors, which can prevent a few inaccurate inputs from dramatically changing the results \citep{lu2020preference}. While these techniques enhance the formal fairness of the aggregation process, they do not fully address the potential for bias introduced by the PRM itself, reiterating the importance of developing a high quality PRM. 

Creating a high quality PRM is nontrivial. \citetessler\space report that the relatively small and dated \textit{Chinchilla}-based reward model (which was used in the HM) was actually more performant than a large, state-of-the-art, out-of-the-box large language model (SM 3.5). The primary reason is that the reward model had been specifically fine-tuned from human preference data for the task. Such a result highlights the importance of careful consideration and curation of training data: ensuring representation across a wide spectrum of demographics, viewpoints, and communication styles will be key to creating fair PRMs. Further, developing tools to help individuals express their preferences more effectively or incorporating feedback loops that allow participants to correct the model's understanding of their views can also help improve reward model accuracy. 

\paragraph{Fair generation}

Even if the processing of counting votes is fair and all voters have an equally easy time expressing their preferences, voters could still feel not well represented by the slate of candidates they have to choose from. The generative model, responsible for creating the candidate group statements, thus plays an important role in the fairness of the system. \citetessler\space did not encode any special mechanism to support a diverse or representative array of candidates, only ensuring that the model did not suggest the exact same candidate multiple times. 

Ideally, though, the generated set of statements should, on average, reflect the diversity of viewpoints proportionally, mirroring the distribution of minority and majority opinions within the group. Candidate statements within that set, however, can and should explore different framings and compromises, perhaps even leaning more heavily towards different perspectives, to allow the PRM to analyze and rank these variations (cf., \citealt{fish2023generative}). 

The embedding analysis used in \citetessler\space could be used to quantify how well the candidate statements capture different perspectives, providing one measure of representativeness on the individual statement level, which could be used to evaluate the representativeness of the set of candidates as a whole. The measure is imperfect though: Common ground often involves novel syntheses and framings that transcend a simple aggregation of initial opinions. Therefore, we encourage more theoretical and empirical work into the quantification of different perspectives vis-a-vis aggregated group statements to better assess fairness in the candidates proposed during AI-mediated deliberation processes.

\subsubsection{Evaluating fairness of Habermas Machine outputs}

Designing the system with the explicit goal of generating statements that fairly represent the underlying opinions of participants is an important development over general-purpose language models, but it is unlikely to be sufficient to guarantee fairness in any formal sense. In general, ensuring AI systems are entirely aligned with the goals set for them remains an open area of research \citep{gabriel2020artificial}. Incorporating AI into democratic processes will therefore necessitate mechanisms for ongoing monitoring and human intervention in cases where misalignment does occur.

For instance, in the context of the HM, the fairness of our social choice aggregation method depends on whether the PRM can accurately rank the candidate statements according to an individual’s preferences \citep{lazar2024can}. While gathering even more training data is likely to further improve the accuracy of the PRM, it is unlikely that any model will ever be able to perfectly predict participant preferences. Moreover, it is difficult to ensure that errors in the model do not systematically bias outputs in subtle ways. For example, if the training data comes from a highly homogeneous group of people, the model may be less accurate for those outside that group, something observed in LLMs \citep{durmus2023towards}. 

To counter this, one key piece of tooling to be explored is to augment the deliberation with real-time visualization and ‘sensemaking’ tools to show the PRM predictions to the people deliberating for them to examine and even correct in real time where the reward model makes an incorrect prediction (e.g., \citealt{jigsaw2024making}). Additionally, as described in the previous section, one could imagine a protocol where the system asks a specific user for more information about their preferences (e.g., clarifying questions) in case the PRM has high uncertainty over its predictions (e.g., \citealt{bakker2019fairness}).

Alongside advanced tooling, it is important to continue to develop analysis and interpretability techniques, such as the embedding analysis used in \citetessler\space which analyzed the influence of majority and minority opinions on the final statement. Similarly, \citetessler\space also evaluated the potential for unfair impacts of bias in HM statements on participants' belief change. This analysis revealed no statistical relationship. These kinds of analysis metrics could be further developed in order to track the behavior of AI mediation systems across developments and deployments. 

\subsection{Participation: Will AI mediation scale deliberation?}

Scalability is a key promise of AI-mediated deliberation, offering the potential to engage far larger and more diverse groups of people in democratic processes. Can the HM effectively facilitate deliberation among increasingly large groups of people? 

In \citetessler, the HM was trained using a large ($n$ = 2110) convenience sample of UK participants from a crowd-sourcing research platform. This sample was not specifically designed to be representative of the UK population. However, analysis of self-reported demographics revealed a reasonable level of representativeness, except for the over-65 age group, which was underrepresented. The HM was reported to perform equally well for a demographically representative sample as for their convenience samples, suggesting some tolerance for demographic misrepresentation in the training procedure. However, targeted evaluations are necessary to confirm this observation and ensure robust performance across diverse population subgroups. For the HM to be used to inform real-world policy making, efforts must be made to encourage and facilitate diverse participation to reflect the target population. This includes everything from targeted recruitment and addressing barriers to participation (e.g., reaching underserved communities, supporting those with low technological literacy) and inclusive design and language (e.g., multi-lingual formats, clear and simple language) to incentives, motivation, and confidentiality (e.g., protecting privacy).

\subsubsection{The challenge of scale in deliberative democracy}

One of the foundational ideals for deliberative democracy is the principle of full inclusion, that “no one capable of making a relevant contribution has been excluded” \citep{habermas2003truth}. Traditional deliberation methods struggle with large-scale participation, owing to the time or cost incurred by participants and organizers of deliberative exercises.\footnote{Given his general opposition to the election of representatives, Jean-Jacques Rousseau argued, for example, that the scale of the ideal republic must be significantly constrained and likely extends to no more than a few thousand people (\citealt{rousseau1893social}, p.88).} What if we didn’t have to make choices about who to include? What if we could include everyone? In \citetessler, small groups (up to five people) were studied, a scale where human mediators are competitive with the AI. If a human had to aggregate twenty, one hundred, or one thousand opinions, the task would not be possible. 

AI mediation tools like the HM greatly expand our vision of scale. The primary technical limitations, in principle, are the digital platform capacity (concurrent users) and the context window size (the amount of text an LLM can consider at one time) of the LLM components. The largest context windows of state-of-the-art LLMs have grown approximately 1,000x (or, three orders of magnitude) between the years 2022 and 2025, and it is conceivable that the context window will stop being a limiting technical parameter in the future. 

\subsection{Scalable oversight of mass deliberations}

Even if an LLM can technically aggregate thousands of opinions, how can we be confident it does so effectively? Simply expanding the HM protocol to large groups and optimizing for endorsement might lead to short, bland statements that say little of substance. While \citetessler\space found that the model produced informative statements in small groups, this may not generalize to much larger, more diverse populations.

At its heart, this is a scalable oversight problem, from the world of AI safety and alignment \citep{amodei2016concrete}. As AI systems become increasingly capable, they may at some point surpass the level of human comprehension \citep{bowman2022measuring}. How are we to verify what the system is telling us in those situations? This is not a futuristic concern: It is already challenging to evaluate a group statement produced by an AI mediator that is aggregating 1,000 opinions (e.g., is the statement based on hallucinated opinions?). As the task becomes more complex and human oversight becomes more unreliable, the risk of rewarding behavior that appears desired but is actually problematic increases \citep{amodei2016concrete, krakovna2020specification}. 

Two approaches from the world of scalable oversight provide possible ways out of this quagmire: AI assistance and task decomposition. Capitalizing on the ability for LLMs to perform arbitrary language-based tasks, one can further train an LLM to issue critiques (natural language critical comments) in a given context. \cite{saunders2022self} showed that assistance from AI models successfully enabled human data labelers to discover more flaws in text summaries, an early demonstration that AI assistance can help humans perform difficult tasks (see also \citealt{jain2025human}). Additionally, providing not just a single critique but rather critiques of critiques or a debate has been proposed as a solution to scalable oversight \citep{brown2023scalable, irving2018ai}, and there is some promising empirical evidence in favor of debate over standalone critique but this is an area of ongoing research \citep{michael2023debate}. This proposal raises intriguing questions about using deliberation to oversee a deliberation assistant.

A second approach to scalable oversight that could yield benefits for supervising mass deliberations is task decomposition, taking a top-level task and decomposing it into several smaller subtasks. Subtasks could be decomposed further until arriving at a task that a human could reliably oversee. \citep{wu2021recursively} explored this method for attempting to faithfully summarize entire books using a fine-tuned version of GPT-3, which could only fit approximately 1500 words at a time into its context window. This was an early demonstration of a hierarchical aggregation method.

\subsection{Hierarchical aggregation: A path to scalable and transparent deliberation}

Hierarchical aggregation could provide additional benefits beyond oversight of a high-quality process. Imagine a group of 1,000 residents deliberating over a local issue. If we split this group into 100 virtual tables each with 10 residents, we can use a similar version of the Habermas Machine as the one tested in \citetessler. After finding common ground at each ‘table,’ we could organize the tables into clusters of 10 tables each (10 clusters of 10 tables). The HM could then try to find agreement among the tables within each cluster. This process could continue a third time to find agreement among the clusters. If the process was successful, we would have found common ground among 1,000 people, but not by directly aggregating all opinions, but by hierarchically aggregating small groups, and then groups of groups. The hierarchical approach may not only be more efficacious, it is more transparent and verifiable, as discussants can inspect statements at various levels of aggregation.

In summary, the Habermas Machine demonstrates considerable promise for scaling deliberative democracy, although important questions remain. While the original training data was limited in representativeness, the evaluation experiments suggested a degree of tolerance, and future efforts should prioritize diverse and inclusive participation. The primary technical barriers to scaling, such as the LLM context window size, are rapidly diminishing. However, scaling introduces the critical challenge of oversight: ensuring the quality and validity of AI-mediated deliberations with large, diverse groups. Promising approaches for scalable oversight include leveraging AI assistance for critique or debate and, perhaps most promisingly, employing task decomposition through hierarchical aggregation. This method, dividing large groups into smaller, interconnected units, offers a path towards both technically feasible and transparently overseeable mass deliberation. Further research and testing are crucial to validate the effectiveness and robustness of these approaches, particularly hierarchical aggregation, in real-world settings. Thus, while the HM holds potential to scale, not just in participant numbers but also in maintaining deliberative integrity, significant empirical investigation is needed to fully realize this potential and address the inherent challenges of large-scale, AI-mediated democratic processes.

\subsection{Deliberation: Can AI enhance deliberative quality?}

Even if many diverse voices are heard and weighed fairly, a deliberative process can falter if the substance of the discussion is poor or ill-informed. “Deliberative quality” is multifaceted, spanning process (e.g., reason-giving; \citealt{steenbergen2003measuring}) and outcome (e.g., intersubjective consistency within groups; \citealt{niemeyer2022deliberative}), with an important component being the extent to which deliberation is grounded in trustworthy, high-quality information \citep{fishkin2018democracy}. The original design of the HM deliberately bracketed this dimension and prioritized the fair aggregation of opinions over an evaluation of their substance \citetesslerp. How can AI systems move beyond merely aggregating opinions to actively enhance the quality of the deliberative inputs themselves?

Efforts to improve deliberative quality are, in fact, central to many existing deliberative democratic practices. For example, citizens’ assemblies (such as Ireland’s Citizens’ Assembly established in 2016) include expert testimony phases drawing on a diverse range of experts in order to ground discussions in evidence \citep{suiter2018deliberation, muller2023reactions}. Deliberative polls provide participants with balanced briefing materials to create a shared, informed foundation \citep{fishkin2005experimenting}. More recent technological innovations offer models for this kind of support at scale, such as the peer-to-peer fact-checking in X’s Community Notes and Taiwan’s crowd-sourced verification tools, both shown to be effective in identifying (and either flagging or removing) low-quality information \citep{martel2024crowds, zhao2023insights}. Do LLMs offer novel ways to enhance the deliberative process?

\subsubsection{Modalities of AI-assisted deliberation}
Large language models present both challenges and opportunities for the epistemic health of public deliberation \citep{weidinger2022taxonomy, mckinney2024integrating, summerfield2025impact}. While factuality remains an ongoing challenge for generative models, recent advances like long context windows (e.g., \citealt{team2024gemini}) and tool use (e.g., being able to search the web) enable them to more effectively attend to the relevant information and answer questions accurately. This technical progress makes the prospect of an AI ‘expert on tap’ increasingly feasible \citep{sprain2014utilizing}, one that could improve the veracity of a deliberation by either drawing from a pre-vetted set of balanced, trustworthy information provided in the AI’s context window or by effectively searching the web for relevant information. This opens up several distinct modalities for how AI can enhance the deliberative process.

At the level of the individual, AI can support participants one-on-one before they even convene as a group (e.g., in an ‘individual deliberation’ phase). This interaction could serve two functions. First, as an epistemic assistant, the AI can help individuals learn about the topic and proposals under discussion, test their assumptions, and, potentially, counter misinformation (e.g., \citealt{costello2024durably}). Second, it could act as a deliberative coach, helping participants clarify and articulate their own perspectives, perhaps helping mitigate against an imbalance between participants based on their verbal ability (“deliberative upskilling”; \citealt{summerfield2025impact}). This could take the form of asking targeted clarifying questions, something the language models can be trained to do effectively \citep{andukuri2024star}.

At the group level, AI can be integrated in several ways. Foremost, an AI could be a useful tool for human mediators. In \citetessler, the HM-generated statements were perceived as more impartial than those written by humans playing the role of mediator. A human facilitator could thus use a tool like the HM to generate draft statements, which they could further edit before sharing with the discussants. In certain settings, where a human mediator is unavailable, the tool could be used in an autonomous mode to act as the mediator itself. A different, bottom-up approach could use AI to facilitate a peer-to-peer process inspired by X’s Community Notes. This model would enable participants to provide context or challenge factual claims made by others in a ‘side-deliberation.’ Here, the AI could help participants find relevant sources for their claims or even synthesize dissenting notes. HM-style aggregation mechanisms could then be used to determine which contextual notes have found broad consensus and should be surfaced to the main group \citep{de2025supernotes}. 

The introduction of any such assistant, regardless of the above-mentioned modalities, raises a set of overlapping challenges. Technically, the AI must provide reliable, near-expert-level information, selecting the most relevant facts from a vast body of knowledge and presenting them in a neutral, accessible manner. From a human-computer interaction perspective, the interaction must be carefully designed to encourage open inquiry and discussion without overwhelming participants, disrupting the flow of conversation, or leading users to predetermined conclusions. There is an ethical tightrope to be walked: there is a delicate balance between improving the quality of deliberation and ‘cooking the books’ by unfairly silencing minority viewpoints or supplanting human agency. Any interaction that builds rapport, such as one-on-one coaching, carries added risks of manipulation that must be carefully monitored \citep{gabriel2024ethics}. The ultimate goal is to create a complementary partnership where the combined human-AI decision process is more robust than either could be alone \citep{steyvers2024three}.

\section{Other Considerations}

\subsection{What does AI-mediated caucus deliberation lack?}

Deliberation can occur in both face-to-face and digital settings, and mediation can take various forms depending on the context. Deliberative technologies like Pol.is, Remesh, and the HM focus on the digital aspect of deliberation, offering a technical solution to the inclusion/participation aspect of Fishkin’s trilemma. One particular design feature of the deliberation protocol used in \citetessler\space is that of caucus mediation, where the participants do not themselves directly interact. This rather absolute form of mediation has the advantage that participants are not influenced by social desirability effects and may be more open to sharing their true feelings. The AI mediator will not judge them for holding a particular view, and their view does not even necessarily need to be attributable to them. 

Despite certain benefits of digital and caucus mediated deliberation, important variables may be lost compared to face-to-face settings \citep{saetra2024computation}. Specifically, fully digital approaches may struggle to foster the positive social connections—trust, empathy, and reciprocity—that often characterize in-person dialogue. Furthermore, the very nature of algorithmic decision-making can trigger algorithmic aversion \citep{dawes1979robust}, wherein individuals distrust or resist outcomes generated by algorithms, even when these outcomes are demonstrably superior to those achieved by humans.

\subsubsection{Social-relational dynamics}

Digital deliberation platforms lack the rich social-relational dynamics of face-to-face interaction. Features like building trust, empathy, and reciprocity with other discussants are largely absent. The impersonal and anonymous nature of caucus mediation, in particular, while beneficial for reducing bias, further limits these social elements (cf., \citealt{restrepo2024softening}).

Deliberations have different goals depending on the context. Conflict resolution, for instance, may prioritize building trust and empathy, while deliberation in business or political settings may primarily be geared towards agreeing on the text of a written document. In face-to-face discussions, social-relational factors likely contribute to reaching agreement. However, in scalable, AI-mediated deliberation, it remains an empirical question whether these factors are essential for producing a high-quality proposal as well as for the long-term acceptance and implementation of that proposal. Successfully translating an in-principle supported policy into a practical solution may be undermined if it does not go hand-in-hand with reciprocal trust between participants.

At the same time, for democratic deliberations that have the potential for real world policy decisions, the outcomes of a deliberation could impact potentially far more people than the discussants who participate in the deliberations \citep{fung2013principle, goodin2007enfranchising}. Thus, we should not only consider the benefits that are conferred to the discussants for in-person vs. AI-mediated digital deliberation, but the costs incurred by those who do not have a seat at the (either physical or digital) table. The ability of digital technologies to scale to very large groups bridges the gap between those involved in the deliberation and those affected by it, an increasingly important dimension to the trade-offs between in-person vs. digital deliberation as AI gets inserted (see also above section on Participation).

\subsubsection{AI-mediated deliberation in its current forms has some positive engagement effects}

Research examining the engagement effects of AI-mediated deliberation is limited at the moment, but some positive signs exist. \cite{gelauff2023achieving} found that deliberations facilitated by an AI mediator on Stanford’s Deliberative Polling platform were rated as high of quality as those mediated by humans along several key dimensions, evidenced by the level of agreement with statements such as “participating in the small group discussion was valuable,” “opposing arguments were considered,” and “I learned a lot about people very different from me.” In \citetessler, post-deliberation surveys (described in the Supplementary Materials) revealed that participants felt they had an impact on the final outcome of the process a majority of the time and almost never felt unwanted pressure to adopt a particular viewpoint. Participants also frequently reported they learned about other viewpoints, learned about the issue more, and that they liked the process. These positive findings suggest that AI mediation, even that which restricts direct interpersonal interaction, can foster a sense of engagement and impact.

\subsubsection{Addressing algorithmic aversion}

Beyond the absence of social-relational dynamics, AI-mediated deliberation—if it lived up to the promise of enhancing participation, equality, and deliberative quality as explored in the above sections—faces the challenge of overcoming algorithmic aversion \citep{dawes1979robust, dietvorst2015algorithm}. In the context of the HM, this aversion could manifest as skepticism towards the impartiality of the generated group statements or resistance to accepting the outcome of the AI-mediated process. This distrust can undermine the legitimacy of the deliberation and hinder the adoption of AI mediation as a valuable tool for democratic decision-making. Mitigating algorithmic aversion requires building trust and transparency in the full deliberation protocol. This can be achieved by providing clear explanations of how the HM functions, emphasizing the system's accordance (both in design and in practice) with democratic principles like fairness and inclusivity. Additionally, one could offer opportunities for participants to understand (e.g. through the aforementioned visualizations) and potentially revise the model's interpretations of their preferences.  Furthermore, demonstrating the HM's effectiveness through empirical evaluations and comparisons with human-mediated deliberations can help to build confidence in its ability to provide positive benefits for deliberation.

\subsection{Where does AI-mediated deliberation fit into the democratic process?}

The analysis so far has focused on whether AI mediation can, in principle, help navigate the democratic trilemma. This theoretical potential, however, only becomes meaningful when grounded in specific applications. Therefore, we now shift our focus from the internal principles of AI mediation to its external role, exploring where systems like the Habermas Machine could be integrated into the broader democratic landscape, from enhancing informal online discussions to augmenting formal policymaking. A prerequisite for exploring these applications is a clear understanding of what systems like the HM are designed to achieve. At its core, its objective is to find “common ground”—a foundational concept whose specific meaning dictates the kinds of democratic problems the system is equipped to help navigate.

\subsubsection{What is common ground?}

The primary objective of the HM is to find "common ground"—a concept rooted in linguistics \citep{stalnaker1975presuppositions} and adapted to deliberative theory \citep{morrissey2023finding} that refers to propositions participants accept for the purposes of conversation without necessarily holding them as deep personal beliefs. This flexible objective can take several forms, from first-person plural statements of shared values (‘We believe ABC’) as seen in \citetessler, to a clarification of disagreement that produces an agenda for a future deliberation (‘The group agrees on X, but disagrees on Y’). By highlighting legitimate areas of difference that the group collectively recognizes \citep{mouffe1999deliberative} AI mediation could achieve an “economization on disagreements” \citep{gutmann2009democracy}, clarifying the points of contention and facilitating subsequent decision-making processes, such as voting or structured negotiation. It could even result in "meta-consensus" where participants accept an outcome as legitimate due to a fair process, even without full agreement on the content of the outcome \citep{bachtiger2018deliberative}. In all cases, the goal is a foundational, shared understanding that can serve as a precursor to subsequent negotiation or decision-making.

Because systems like the HM are based on generative language models, their role is highly customizable and extends beyond simply finding common ground. An AI may help in restating individual opinions to help promote mutual understanding (cf., \citealt{argyle2023leveraging}), issuing clarifying questions to probe areas of disagreement, or proposing concrete action plans, even perhaps provocatively if no participant had yet proposed specific plans. The potential for the latter functionality (i.e., proposing new courses of action) brings into relief that the system is not intended to produce binding policies on its own, but rather to provide fodder to catalyze further discussion and action: The AI's outputs serve as proposals, subject to critique, refinement, and ultimately, approval or rejection by the human participants. It is this iterative exchange—AI proposal, human feedback, AI revision—that holds the potential to move deliberations forward productively. In any deliberative application, the role of AI should be to support discussion, not to collapse opinions into a forced perspective or to manipulate participants into agreement.

\subsubsection{Potential applications of AI mediation in democratic settings}

The range of possible applications of the core technology in democratic settings is wide and sometimes the objective may extend beyond finding common ground in order to reach a decision or plan of action. Here, we describe a few potential applications: enhancing informal democratic processes via the digital public square, scaling existing deliberative initiatives (e.g., deliberative polls), and, more speculatively, as a mechanism to negotiate over the shared text in a multilateral negotiation.

One very clear application of AI mediation to democracy is in the informal setting, the digital public square (i.e., the internet; \citealt{rheingold1993virtual, zuckerman2013digital, goldberg2024ai}). Active AI mediation could elevate these digital spaces by enabling constructive interaction between participants \citep{argyle2023leveraging}. Community-based approaches to fact-checking such as X’s Community Notes could be plausible application settings for AI-mediated deliberation. There is some early promising evidence that a tool similar in spirit to the HM can be used to improve the quality of community notes \citep{de2025supernotes}. Short of being deployed on large social media platforms, the HM could be embedded with a Collective Dialogue System, a tool that enables (potentially large) groups of participants to express themselves and interact with others, with an eye towards synthesis and understanding of the collective body \citep{konya2023democratic, ovadya2023reimagining}. Collective Dialogue Systems can play an important role in helping the public set an agenda for more formal political deliberations (e.g., citizens’ assemblies, or crafting a political party platform). They may also be used as an alternative to traditional polling to get a richer sense of what a population thinks. 

AI mediation could additionally help scale existing deliberative initiatives, such as deliberative polling. A deliberative poll is a method of gauging public opinion by gathering a representative sample of participants, providing them with information on an issue, and facilitating structured discussions before re-measuring their views. Deliberative polls have the goal of both understanding what a population thinks about an issue (when given time to consider and weigh different perspectives) as well as to educate the participants about the issue further (through the free and equal exchange of views). This latter goal in particular, the educative benefits, are only conferred to the participants who are involved in the deliberation, thus limiting this technique as a tool for education of the electorate. Recent findings suggest similar educative benefits can be conferred through a video chat-based deliberation, potentially enabling the scaling of these processes \citep{gelauff2023achieving}. With AI mediation, the cost of mediation would come down significantly and the process could be further scaled to bring in an even broader group of participants. This might further spur the adoption of deliberative exercises and create more public will around incorporating deliberative democracy elements into the formal democratic process. Note, however, that the mediation in a deliberative poll is primarily focused on real-time discourse management and making sure everyone has a chance to speak. Introducing the HM would create an opportunity for the creation of new policy statements or positions, which would constitute an innovation on the deliberative poll.

More speculatively, AI mediation could be used to facilitate multilateral negotiations over the text of a document. Several problems take this form: aligning diverse stakeholders to develop a political party platform, different political parties debating over the wording of legislation, or even multiple nation-states coming to an agreement on international regulations. Of course, the different stakeholders involved would have to agree on the use of AI prior to engaging in the negotiation process. Additionally, AI mediation might only be used for part of the negotiation or as the first phase of a negotiation, with details of higher-stakes agreements left to existing negotiation practices.

The applications of AI mediation are not confined to state-level democratic issues. In fact, some of the most practical and immediate uses may lie in addressing local issues within quasi-democratic settings. Contexts such as university governance, neighborhood associations, school boards, and non-profit governance frequently require reconciling diverse perspectives to reach agreement. While not traditional democracies, these organizations often employ democratic tools like voting or consensus-building. AI mediation could therefore serve as a valuable new instrument in these environments, streamlining decision-making processes and helping to ensure all voices are heard. 

\section{Discussion}

Many decision-making bodies, from democratic governments to quasi-democratic organizations like university governance boards and neighborhood associations, face challenges in enacting policies that reflect the interests and values of citizens and members simply due to a lack of information about the spectrum of viewpoints. In democracies, not all issues are equally salient in electoral competition, and coalitions do not form naturally around existing party affiliations for every issue. Public opinion can also be misrepresented when interested parties have unequal access to resources. The tools available to governments and decision-makers to proactively surface additional input from citizen stakeholders, such as public consultations and opinion polling, have traditionally been limited in scale, depth, or to quantitative rather than qualitative measures. In this paper, we have outlined the potential for AI-powered systems like the HM to deepen public participation in decision-making processes (see also \citealt{birhane2022power}). In principle, these systems can be designed and built with the ability to foster inclusivity, and to surface informed and meaningful contributions at scale.

The work of \citetessler\space provided empirical evidence that a purpose-built AI system, the HM, can effectively mediate common ground. In this paper, we have used this work as a lens to explore how AI might help navigate Fishkin's democratic trilemma. We examined how AI can address each component: upholding political equality through fairer mediation; enhancing participation through scalability; and improving the quality of deliberation with tools for surfacing trustworthy information and assisting participants. Much work remains to realize this potential. Exploring different interaction models beyond strict caucus mediation, or further refining the integration of expert testimony and deliberative assistants, are important for future research and development. Continued focus on these dimensions is necessary for the safe and effective deployment of such systems in real-world democratic processes.

Yet other issues to consider remain before deploying such systems in practice. How can the security of an AI mediator system be effectively safeguarded during both training and deployment, and how can the privacy of participants who contribute potentially sensitive opinions to the system be protected? Who should be responsible for training and deploying the system?

Another important consideration is how strategic actions by participants could impact the deliberative outcomes of an AI mediated process. While the HM strives for a balanced representation of viewpoints, it assumes that individuals sincerely input their preferences. If participants strategically misrepresent their opinions, then the resulting group statements may not accurately reflect the true common ground, potentially skewing the deliberation towards a suboptimal outcome. Strategic voting can sometimes encourage coalition and consensus-building in specific voting contexts, but the complexity of AI systems (including e.g., the HM) makes predicting the effects of such behavior challenging. Even without a precise theoretical understanding of the system, some participants may gain an understanding of how to manipulate the result after repeated interactions with the HM \citep{holliday2025learning}. Further research and red teaming efforts are hence needed to investigate what constitutes rational strategic behavior in the context of AI-mediated deliberation, the likelihood of such misrepresentations, including exploring methods for detecting insincere inputs, for designing mechanisms to encourage truthful participation (e.g., \citealt{flanigan2023distortion}), and potentially for building an assumption of strategic behavior into the aggregation component of the system. 

There is also a large set of open questions regarding how the technology would operate in the real world. Would participants engage in deliberation in good faith in the presence of genuine political stakes? Would participation in these deliberations genuinely enhance subjective perceptions of democratic representativeness? Can enough trust be established in the HM that it can withstand political contestation?

The convergence of AI and democratic deliberation offers a potent avenue for enhancing citizen engagement and fostering more representative governance. While systems like the Habermas Machine are not a panacea for all democratic challenges, they can represent a significant stride towards augmenting traditional deliberative processes. Further research and development, guided by the principles of fairness, transparency, and inclusivity, promise to unlock the full potential of AI-mediated deliberation, creating a future where technology empowers citizens to shape the policies that affect their lives.

\bibliography{main}

\end{document}